\documentclass
[superscriptaddress,secnumarabic,amssymb,amsmath,nobibnotes,aps,prd,showkeys,showpacs,nofootinbib,onecolumn,12pt]{revtex4}%
\usepackage{graphicx}
\usepackage{epsf}
\usepackage{bm}
\usepackage{amsmath}
\usepackage{amsfonts}
\usepackage{amssymb}%
\usepackage{color}

\providecommand{\U}[1]{\protect\rule{.1in}{.1in}}

\newcommand{\be}{\begin{equation}}
\newcommand{\ee}{\end{equation}}

\newcommand{\mincir}{\raise
-3.truept\hbox{\rlap{\hbox{$\sim$}}\raise4.truept\hbox{$<$}\ }}
\newcommand{\magcir}{\raise
-3.truept\hbox{\rlap{\hbox{$\sim$}}\raise4.truept\hbox{$>$}\ }}

\usepackage{tikz,xcolor,hyperref}

\definecolor{lime}{HTML}{A6CE39}
\DeclareRobustCommand{\orcidicon}{%
	\begin{tikzpicture}
	\draw[lime, fill=lime] (0,0) 
	circle [radius=0.16] 
	node[white] {{\fontfamily{qag}\selectfont \tiny ID}};
	\draw[white, fill=white] (-0.0625,0.095) 
	circle [radius=0.007];
	\end{tikzpicture}
	\hspace{-2mm}
}

\foreach \x in {A, ..., Z}{%
	\expandafter\xdef\csname orcid\x\endcsname{\noexpand\href{https://orcid.org/\csname orcidauthor\x\endcsname}{\noexpand\orcidicon}}
}

\begin{document}
\title{Multiscalar-torsion Cosmology: Exact and Analytic Solutions from Noether Symmetries}
\author{K. Dialektopoulos\orcidA{}}
\email{kdialekt@gmail.com}
\affiliation{Department of Physics, Nazarbayev University, 53 Kabanbay Batyr avenue, 010000 Astana, Kazakhstan}
\affiliation{Laboratory of Physics, Faculty of Engineering, Aristotle University of Thessaloniki, 54124 Thessaloniki, Greece}
\author{G. Leon \orcidB{}}
\email{genly.leon@ucn.cl}
\affiliation{Departamento de Matem\'{a}ticas, Universidad Cat\'{o}lica del Norte, Avda.
Angamos 0610, Casilla 1280 Antofagasta, Chile}
\affiliation{Institute of Systems Science, Durban University of Technology, PO Box 1334,
Durban 4000, South Africa}
\author{A. Paliathanasis  \orcidC{}}
\email{anpaliat@phys.uoa.gr}
\affiliation{Institute of Systems Science, Durban University of Technology, PO Box 1334,
Durban 4000, South Africa}
\affiliation{Departamento de Matem\'{a}ticas, Universidad Cat\'{o}lica del Norte, Avda.
Angamos 0610, Casilla 1280 Antofagasta, Chile}

\begin{abstract}
The Noether symmetry analysis is applied in a multi-scalar field cosmological model in teleparallel gravity. In particular, we consider two scalar fields with interaction in scalar-torsion theory. The field equations have a minisuperspace description, and the evolution of the physical variables depends on the potential function that drives the scalar fields' dynamics.
With the requirement for the field equations to admit non-trivial Noether point symmetries and the use of the first theorem of Noether, we constraint all the functional forms for the potential. Finally, we apply symmetry vectors and the corresponding conservation laws to determine exact and analytic solutions in multiscalar-torsion cosmology.

\end{abstract}
\date{\today}
\maketitle

\section{Introduction}

\label{sec1}

While General Relativity is a well-tested theory of gravity, it cannot provide
a physical mechanism to the problems which follow from cosmological
observations \cite{obs1,obs2}. Nowadays, cosmologists deal with alternative
and modified theories of gravity by keeping the geometric character of the
gravitational theory \cite{vv1,vv3}. With the term teleparallelism \cite{rev1}
we describe a family of gravitational theories where the fundamental geometric
object is the teleparallel connection \cite{Weitzenb23} related to the
nonholonomic frame, in contrast to General Relativity, where the fundamental
the geometric object is the Levi-Civita connection related to the metric tensor.

As in the case of General Relativity, in which the gravitational Lagrangian is defined by the Ricci scalar $R$ of the Levi-Civita connection, in teleparallel geometry, the torsion scalar $T$ is considered for the definition of Action Integral. When the gravitational Action Integral is linear to the torsion
scalar $T$, the theory is equivalent to general relativity, and it is known as the
Teleparallel Equivalent of General Relativity (TEGR) \cite{ein28, Hayashi79}.
However, the equivalence ends here because when one introduces the gravitational action, scalar fields that are coupled to
gravity \cite{te1,te2,te3,ss3} or nonlinear terms of the
geometric scalars \cite{ft5,ft6,st1,st2}, the two approaches give different results. The latter
is easy to understand because the Ricci scalar includes second-order
derivatives of the metric tensor, while the torsion scalar $T$ includes only first-order derivatives for the arbitrary functions of the nonholonomic frame.

On the other hand, scalar fields play a significant role in the description of
the matter component responsible for the different phases of the universe
during the cosmic evolution \cite{sf1,sf2}. Indeed, the early accelerated era
of the universe known as inflation is attributed to the inflaton scalar field
\cite{inf1,inf2,inf3}. Furthermore, scalar fields have been used as 
unified dark energy models that can describe the late-time acceleration phase
of the universe and the dark matter component \cite{uni1,uni2,uni3}. 
Scalar fields provide a simple mathematical mechanism for the introduction of
additional degrees of freedom in the field equations, consequently enriching the cosmological dynamics and evolution. Thus, multiscalar field
models in gravity have been widely considered by cosmologists over the last
years; see, for instance, \cite{mf1,mf2,mf3,mf4,mf5,mf6} and references therein.

In this study, we are interested in a multiscalar field cosmological model in
the context of scalar-torsion theory. Scalar-torsion theory is the analogue of
the scalar-tensor theory in teleparallelism in which the scalar field is
non-minimally coupled to gravity with interaction in the Action Integral
between the scalar field and the torsion scalar $T$ \cite{te1,te2,te3}.
Another important characteristic of the scalar-torsion theory is that the
scalar field can attribute the additional degrees of freedom to\ a
higher-order teleparallel theory \cite{ss3}, similar to Horndeski gravity 
\cite{Bahamonde:2019shr} or to the
relation of the O'Hanlon theory with $f\left(  R\right)  $-gravity \cite{vv3}.
Furthermore, we consider the existence of a second scalar field minimally
coupled to gravity but with an interacting term in the kinetic part with the
scalar-torsion field. Specifically, we define a model similar to that studied
before in scalar-tensor theory in \cite{ini2}. This family of models can
provide the hyperbolic inflation epoch \cite{mf2} in the Jordan frame
\cite{ini3}. For the background space, we assume the spatially flat
Friedmann--Lema\^{\i}tre--Robertson--Walker (FLRW) metric, which admits six
isometries. We consider the scalar fields to inherit the symmetries of the
background space from where it follows that the gravitational field equations
are reduced into ordinary differential equations. Moreover, the study the
evolution and the asymptotic dynamics of the physical parameters has been widely
applied in many physical theories to construct criteria that a
proposed theory should satisfy to be cosmologically viable
\cite{dn1,dn2,dn3,dn4,dn5,dn6,dn7}. Furthermore, the Noether symmetry analysis
is considered to determine conservation laws for the field equations
and infer the integrability properties. Conservation laws are applied
for the construction of analytic solutions. 

The structure of the paper is as
follows: in Section \ref{sec2}, we present the cosmological model of our consideration
which is a two-scalar field model in the context of teleparallelism.
Specifically, we consider the scalar-torsion theory with a second scalar field
minimally coupled to gravity but with an interaction between the two scalar
fields. We assume that the interaction between the two scalar fields is
provided by the potential function and by the coupling function of the
scalar-torsion model. The field equations form a Hamiltonian system with a minisuperspace description and a point-like Lagrangian exist. Hence,
Noether's theorem is applied in Section \ref{sec3} to constrain the
unknown parameters of this multi-scalar field model. We find seven families of
potential functions where non-trivial conservation laws exist. In Section
\ref{sec4}, we use the Noether symmetry classification to determine
exact solutions and Liouville integrable cosmological models to derive analytic solutions in a multi-scalar field
cosmological model in the teleparallel theory of gravity. The qualitative
behaviour of the solutions is discussed. Finally, in Section \ref{conc}, we
summarize our results and draw conclusions.

\section{Multiscalar-torsion cosmology}

\label{sec2}

We consider the gravitational Action integral in teleparallelism%
\begin{equation}
S=\frac{1}{16\pi G}\int {\rm d}^{4}xe\left[  F\left(  \phi\right)  \left(
T+\frac{\omega}{2}\phi_{;\mu}  \phi^{\mu}  +V\left(  \phi  \right)
+L\left(  x^{\kappa},\psi,\psi_{;\mu}  \right)  \right)  \right]. \label{d.08}%
\end{equation}
where $T$ is the torsion scalar defined by the antisymmetric Weitzenb\"{o}ck
connection $\hat{\Gamma}^{\lambda}{}_{\mu\nu}$, that is,
\[
T={S_{\beta}}^{\mu\nu}{T^{\beta}}_{\mu\nu},
\]
in which $T_{\mu\nu}^{\beta}$ is the torsion tensor defined as $T_{\mu\nu
}^{\beta}=\hat{\Gamma}_{\nu\mu}^{\beta}-\hat{\Gamma}_{\mu\nu}^{\beta}$ and
${S_{\beta}}^{\mu\nu}=\frac{1}{2}({K^{\mu\nu}}_{\beta}+\delta_{\beta}^{\mu
}{T^{\theta\nu}}_{\theta}-\delta_{\beta}^{\nu}{T^{\theta\mu}}_{\theta})$
$\ $where now ${K^{\mu\nu}}_{\beta}=-\frac{1}{2}({T^{\mu\nu}}_{\beta}%
-{T^{\nu\mu}}_{\beta}-{T_{\beta}}^{\mu\nu})~$is the contorsion tensor. We
remark that the Weitzenb\"{o}ck connection $\hat{\Gamma}^{\lambda}{}_{\mu\nu}$
is related to the vierbein fields $e_{i}=h_{i}^{\mu}
\partial_{i}$, as $\hat{\Gamma}^{\lambda}{}_{\mu\nu}=h_{a}^{\lambda}%
\partial_{\mu}h_{\nu}^{a}$ with the metric tensor $g_{\mu\nu}  ~$to be defined as $g_{\mu\nu}=\eta_{ij}h_{\mu
}^{i}h_{\nu}^{j}$. As mentioned above, we work with the Weitzenb\"ock connection instead of the teleparallel one, where the spin connection $\omega ^a{}_{b\mu}$ vanishes.

The Lagrangian density $L\left(  x^{\kappa},\psi
,\psi_{;\mu}  \right)  $ is considered to describe the dynamics of a
second-scalar field, that is,
\begin{equation}
L\left(  x^{\kappa},\psi,\psi_{;\mu}  \right)  =\frac{\beta}{2}\psi_{;\mu}\left(  x^{\kappa
}\right)  \psi^{;\mu}  +U\left(  \psi  \right).
\end{equation}
Hence, the gravitational Action Integral (\ref{d.08}) reads%
\begin{align}
S & =\frac{1}{16\pi G}\int {\rm d}^{4}xe\Big[  F\left(  \phi\right)  \left(
T+\frac{\omega}{2}\phi_{;\mu}  \phi^{\mu}  +V\left(  \phi  \right)
+\frac{\beta}{2}\psi_{;\mu}  \psi^{;\mu}  \right)  +\hat{V}\left(  \phi
,\psi  \right)  \Big]\,, \label{dd.01}%
\end{align}
where the potential function $\hat{V}\left(  \phi
,\psi  \right)  $ has been introduced to describe the interaction between
the two scalar fields. $U(\psi)$ has also been considered vanishing.

At large scales, the universe is assumed to be described by the spatially
flat FLRW geometry with line element%
\begin{equation}
{\rm d}s^{2}=-N^{2}\left(  t\right)  {\rm d}t^{2}+a^{2}(t)({\rm d}x^{2}+{\rm d}y^{2}+{\rm d}z^{2}).
\label{d.06}%
\end{equation}
where $a\left(  t\right)  $ is the scale factor and $N\left(  t\right)  $ is
the lapse function.

A proper set of vierbein fields where the limit of General Relativity is
recovered for $\phi  =$ constant and $\psi =$ constant, is the following%
\begin{equation}
h_{~\mu}^{i}(t)=\mathrm{diag}(N\left(  t\right),a(t),a(t),a(t)),
\end{equation}
from which we calculate the torsion scalar $T=6H^{2},$ with $H=\frac{1}%
{N}\frac{\dot{a}}{a}$, $\dot{a}=\frac{{\rm d}a}{{\rm d}t}$, to be the Hubble function.

By replacing the latter expression for the torsion scalar in the Action
Integral (\ref{dd.01}) and assume that the scalar fields inherit the
isometries of the FLRW universe, we derive the point-like Lagrangian%
\begin{equation}
\mathcal{L}\left(  N,a,\dot{a},\phi,\dot{\phi},\psi,\dot{\psi}\right)
=\frac{F\left(  \phi\right)  }{N}\left(  6a\dot{a}^{2}+a^{3}\left(
\frac{\omega}{2}\dot{\phi}^{2}+\frac{\beta}{2}\dot{\psi}^{2}\right)
+N^{2}a^{3}V\left(  \phi,\psi\right)  \right).  \label{dd.03}%
\end{equation}
where $\hat{V}\left(  \phi,\psi\right)  =V\left(  \phi,\psi\right)  F\left(
\phi\right). ~$

Without loss of generality we select $N\left(  t\right)  =1$ and the
corresponding gravitational field equations read%
\begin{equation}
6H^{2}+\frac{\omega}{2}\dot{\phi}^{2}+\frac{\beta}{2}\dot{\psi}^{2}-V\left(
\phi,\psi\right)  =0~, \label{dd.02}%
\end{equation}%
\begin{equation}
\dot{H}+\frac{3}{2}H^{2}-\left(  \frac{1}{4}\left(  \frac{\omega}{2}\dot{\phi
}^{2}+\frac{\beta}{2}\dot{\psi}^{2}+V\left(  \phi\right)  \right)  -H\dot
{\phi}\left(  \ln\left(  F\left(  \phi\right)  \right)  \right)  _{,\phi
}\right)  =0~,
\end{equation}%
\begin{equation}
\ddot{\phi}+3H\dot{\phi}+\frac{1}{\omega}\left(  \ln\left(  F\left(
\phi\right)  \right)  \right)  _{,\phi}\left(  \frac{\omega}{2}\left(
\ln\left(  F\left(  \phi\right)  \right)  \right)  _{,\phi}\dot{\phi}%
^{2}-\frac{\beta}{2}\dot{\psi}^{2}-6H^{2}-V\left(  \phi,\psi\right)  \right)
-\frac{1}{\omega}V_{,\phi}=0~,
\end{equation}
and
\begin{equation}
\ddot{\psi}+3H\dot{\psi}+\left(  \ln\left(  F\left(  \phi\right)  \right)
\right)  _{,\phi}\dot{\phi}\dot{\psi}-\frac{1}{\beta}V_{,\psi}=0.
\end{equation}

The dynamical system of second-order ordinary differential equations is
autonomous and admits as conservation law the constraint equation
(\ref{dd.02}) which can be seen as the Hamiltonian function%
\begin{equation}
\mathcal{H}\equiv F\left(  \phi\right)  \left(  6a\dot{a}^{2}+a^{3}\left(
\frac{\omega}{2}\dot{\phi}^{2}+\frac{\beta}{2}\dot{\psi}^{2}\right)
-a^{3}V\left(  \phi,\psi\right)  \right)  =0. \label{dd.04}%
\end{equation}

The evolution of the cosmological variables depends on the nature of the scalar
field potential $V\left(  \phi,\psi\right)$. The scope of this work is to
define the functional forms of $V\left(  \phi,\psi\right)  $ by using the
Noether symmetry analysis. Specifically, we shall perform a classification of
the potential function $V\left(  \phi,\psi\right)  $ such that the point-like
Lagrangian to admit non-trivial Noether symmetries, that is, non-trivial
conservation laws. Such analysis has been the subject of study in various
cosmological models.

The novelty of this approach is two-fold. The determination of conservation
laws for the field equations are essential for the study of the integrability
properties of a given theory or for the construction of invariant functions
which can describe the dynamical evolution of the physical explicitly
variables in a specific region in the space of solutions. On the other hand,
Noether symmetries of the field equations are related to the collineations
of the minisuperspace; that is, it is the geometry of the minisuperspace which
impose the existence of Noether symmetries and the specific forms of the
potential function $V\left(  \phi,\psi\right)  $. Hence, the Noether symmetry analysis is a geometric approach to the analysis of gravitational models.

\section{Noether symmetry analysis}

\label{sec3}

Let us briefly discuss the symmetries of differential equations and present
the Noether's two theorems for one-parameter point transformations.

Consider a system of second-order differential equations%
\begin{equation}
\ddot{y}^{A}=\omega^{A}\left(  t,y^{B},\dot{y}^{B}\right), \label{ll.01}%
\end{equation}
where $t$ is the independent variable and $y^{A}$ are the dependent variables;
in our cosmological model $y^{A}=\left(  a,\phi,\psi\right)  $.

Let the function $\Phi$ be the map of one parameter point transformations $\Phi
$\thinspace$:\left\{  t,y^{B}\right\}  \rightarrow\left\{  \bar{t}\left(
t,y^{B},\varepsilon\right),\bar{y}^{B^{\prime}}\left(  t,y^{B}%
,\varepsilon\right)  \right\}  $, where $\varepsilon$ is an infinitesimal
parameter. Function $\Phi\,$\ maps solutions of the system (\ref{ll.01}) into
solutions if and only if%
\begin{equation}
X^{\left[  2\right]  }\left(  \ddot{y}^{A}-\omega^{A}\left(  t,y^{B},\dot
{y}^{B}\right)  \right)  =0~, \label{ll.02}%
\end{equation}
where $X^{\left[  2\right]  }$ is the second extension of the vector field
\begin{equation}
X=\xi\left(  t,y^{B},\varepsilon\right)  \Bigg|_{\varepsilon=0}\partial
_{t}+\eta^{A}\left(  t,y^{B},\varepsilon\right)  \Bigg|_{\varepsilon
=0}\partial_{A}%
\end{equation}
with
\begin{equation}
\xi\left(  t,y^{B},\varepsilon\right)  =\frac{\partial\bar{t}}{\partial
\varepsilon},~\eta^{A}\left(  t,y^{B},\varepsilon\right)  =\frac{\partial
\bar{y}^{A}}{\partial\varepsilon}.
\end{equation}
Therefore, the second extension is given by $X^{\left[  2\right]  }%
=X+\eta^{\left[  1\right]  A}\partial_{A}+\eta^{\left[  2\right]  A}%
\partial_{A}$, in which $\eta^{\left[  1\right]  A}=\dot{\eta}^{A}-\dot{y}%
^{A}\dot{\xi}$ and $\eta^{\left[  2\right]  A}=\dot{\eta}^{\left[  1\right]
A}-\ddot{y}^{A}\dot{\xi}$. Finally, when the symmetry condition (\ref{ll.02})
is true, the generator $X$ will be called a Lie point symmetry for the
dynamical system (\ref{ll.01}).

For dynamical systems which follow from a variational principle, Emmy Noether
published in a pioneer work two theorems which relate the symmetries of the
differential equations to the variational symmetries and to the existence of
invariant functions, which are conservation laws.

Consider now the Action Integral $S=\int\mathcal{L}\left(  t,y^{A},\dot{y}%
^{A}\right)  {\rm d}t$ which describes the dynamical system (\ref{ll.01}), where
$\mathcal{L}\left(  t,y^{A},\dot{y}^{A}\right)  $ is the Lagrange function.
Noether's first theorem states that a Lie symmetry for the dynamical system
(\ref{ll.01}) is also a variational symmetry for the Action Integral $S$, if
and only if there exist a function $f\left(  t,y^{A},...\right)  $ such that
the following condition to be true%
\begin{equation}
X^{\left[  1\right]  }\mathcal{L}+\mathcal{L}\dot{\xi}=\dot{f}.
\end{equation}
If the latter condition is true, the vector field $X$ is characterized as
Noether symmetry. Function $f$ is a boundary term introduced to allow for the
infinitesimal changes in the value of the Action Integral produced by
an infinitesimal change in the boundary of the domain caused by the
transformation of the variables in the Action Integral.

Noether's second theorem provides a simple and systematic way for the
construction of conservation laws for each Noether symmetry. Indeed, if $X$ is
a given Noether symmetry for the dynamical system (\ref{ll.01}) described by
the Lagrange function $\mathcal{L}\left(  t,y^{A},\dot{y}^{A}\right)  $, then,
quantity%
\begin{equation}
I=\left(  \dot{y}^{A}\frac{\partial\mathcal{L}}{\partial\dot{y}^{A}%
}-\mathcal{L}\right)  \xi-\frac{\partial\mathcal{L}}{\partial\dot{y}^{A}}%
\eta^{A}+f. 
\end{equation}
is a conservation law, that is $\dot{I}=0$.

For dynamical systems described by Lagrangian functions of the form%
\begin{equation}
\mathcal{L}\left(  t,y^{A},\dot{y}^{A}\right)  =\frac{1}{2}\gamma_{AB}\left(
y^{C}\right)  \dot{y}^{A}\dot{y}^{B}-U\left(  y^{C}\right),
\end{equation}
it has been found that Noether symmetries are constructed by the elements of
the Homothetic algebra of the second-rank tensor, i.e. the metric tensor,
$\gamma_{AB}\left(  y^{C}\right)  $.

Lagrangian function (\ref{dd.03}) consists of two unknown functions, function
$F\left(  \phi\right)  $, which defines the geometry of the minisuperspace and
the effective potential $V_{\rm eff}\left(  a,\phi,\psi\right)  =a^{3}F\left(
\phi\right)  V\left(  \phi,\psi\left(  x^{\kappa
}\right)  \right)  $. Hence, by following the analysis described in
\cite{Tsamparlis:2018nyo}, the Noether symmetry analysis is two-fold: firstly, we shall
classify the functional forms of $F\left(  \phi\right)  $ where the
minisuperspace of (\ref{dd.03}) admits Homothetic vector fields, secondly, the
homothetic vectors will be used to constraint the functional form of the
effective potential $V_{\rm eff}\left(  \phi,\psi\right)  $ and write the
corresponding Noether symmetry and conservation law.

\subsection{Noether symmetry classification}

Consider now that $F\left(  \phi\right)  $ is a non-constant function. Then,
the classification of the Homothetic algebra for the minisuperspace gives
three cases, $F_{A}\left(  \phi\right)  $ is arbitrary, $F_{B}\left(
\phi\right)  =F_{0}e^{2K\phi},$ where $K$ is an arbitrary non-zero constant;
and $F_{C}\left(  \phi\right)  =F_{0}e^{K\phi}$, with $K=\pm
\frac{\sqrt{3\left(  -\omega\right)  }}{2}$.

For $F_{A}\left(  \phi\right)  $ the Homothetic algebra of the minisuperspace
has two dimensions, and it consists of the Killing vector%
\begin{equation}
K^{1}=\partial_{\psi},
\end{equation}
and the proper Homothetic vector%
\begin{equation}\label{eq:HomVec}
Y=\frac{2}{3}a\partial_{a}.
\end{equation}
On the other hand, for functions $F_{B}\left(  \phi\right)  $ and
$F_{C}\left(  \phi\right)  $ the Homothetic algebras admitted by the
minisuperspace are of four and five dimensions, respectively. Indeed, for
$F_{B}\left(  \phi\right)  $ the additional Killing vector fields are
\begin{equation}
K^{2}=-\frac{2}{3}Ka\psi\partial_{a}+\psi\partial_{\phi}+\frac{8K\ln
a-\omega\phi}{\beta}\partial_{\psi}~,
\end{equation}
and
\begin{equation}
K^{3}=-\frac{2}{3}Ka\partial_{a}+\partial_{\phi}.
\end{equation}

Furthermore, when $K=+\frac{\sqrt{3\left(  -\omega\right)  }}{4}$, that is, in
the case $F_{C}\left(  \phi\right)  $, the fourth Killing vector field of the
minisuperspace is
\begin{equation}
K^{4}=\left(  -\frac{2}{3}a-\frac{a\sqrt{-3\omega}}{6}\phi+a\ln a\right)
\partial_{a}+\left(  \phi-\frac{6}{\sqrt{-3\omega}}\ln a\right)
\partial_{\phi}+\psi\partial_{\psi}\text{.}%
\end{equation}
At this point we remark that the case $K=-\frac{\sqrt{3\left(  -\omega\right)
}}{2}$ is recovered under the change of variables $\phi\rightarrow-\phi$.

As far as the classification of the potential function $V\left(  \phi,\psi  \right)  $ is concerned for
each case of the coupling function $F\left(  \phi\right)  $ follows.

For arbitrary function $F_{A}\left(  \phi\right)  $, and arbitrary potential
there exist the trivial Noether symmetry $\partial_{t}$ with corresponding
conservation law the constraint condition (\ref{dd.02}).

For $V_{1}\left(  \phi,\psi\right)  =V\left(  \phi\right)  $, the field
equations admit the Noether symmetry vector $K^{1}$ with corresponding
conservation law
\begin{equation}
I^{1}\left(  K^{1}\right)  =a^{3}e^{2K\phi}\beta\dot{\psi}.
\end{equation}
Moreover, for $V_{2}\left(  \phi,\psi\right)  =e^{-4\delta\psi}V\left(
\phi\right)  $, there exists the Noether symmetry $\delta\left(  2t\partial
_{t}+Y\right)  +K^{1}$, with conservation law%
\begin{equation}
I^{2}\left(  \delta\left(  2t\partial_{t}+Y\right)  +K^{1}\right)  =2\delta
t\mathcal{H}-e^{2K\phi}a^{2}\left(  8\delta\dot{a}+\beta a\dot{\psi}\right). 
\end{equation}

For the exponential function $F_{B}\left(  \phi\right)  $ there exist
additional functional forms of the potential where Noether symmetries exist.
Indeed, when $V_{3}\left(  \phi,\psi\right)  =V\left(  \psi\right)  $ the
vector field $K^{3}$ is a Noether symmetry with conservation law%
\begin{equation}
I^{3}\left(  K^{3}\right)  =-e^{2K\phi}a^{2}\left(  8K\dot{a}-\omega
a\dot{\phi}\right). 
\end{equation}
When $V_{4}\left(  \phi,\psi\right)  =V\left(  \psi-\frac{\phi}{\alpha
}\right)  $, there exist the Noether symmetry $K^{1}+\alpha K^{3}$ with
conservation law%
\begin{equation}
I^{4}\left(  \alpha K^{1}+K^{3}\right)  =I^{1}\left(  K^{1}\right)  +\frac
{1}{\alpha}I^{3}\left(  K^{3}\right). 
\end{equation}
Furthermore, for $V_{5}\left(  \phi,\psi\right)  =V\left(  \alpha\psi
-\phi\right)  e^{-\bar{\delta}\psi},~\bar{\delta}=\frac{\delta}{a}$, there
exist the additional Noether symmetry is the vector field $\delta\left(
2t\partial_{t}+Y\right)  +K^{1}+\alpha K^{3}$ where now the Noether
conservation law is%
\begin{equation}
I^{5}\left(  \delta\left(  2t\partial_{t}+Y\right)  +K^{1}+\alpha
K^{3}\right)  =I^{2}\left(  \delta\left(  2t\partial_{t}+Y\right)
+K^{1}\right)  +\alpha I^{3}\left(  K^{3}\right)  \text{.}%
\end{equation}
For $V_{6}\left(  \phi,\psi\right)  =V_{0}$, there exist the additional
Noether symmetry $K^{2}$ with corresponding conservation law%
\begin{equation}
I^{6}\left(  K^{2}\right)  =e^{2K\phi}a^{2}\left(  -8Ka\psi\dot{a}+\omega
a\psi\dot{\phi}+a\left(  8K\ln a-\omega\phi\right)  \dot{\psi}\right)
\text{.}%
\end{equation}
When $V_{7}\left(  \phi,\psi\right)  =V\left(  \psi\right)  e^{-4\delta\phi
},~$ the vector field $\delta\left(  2t\partial_{t}+Y\right)  +K^{3}$ is a
Noether symmetry with conservation law%
\[
I^{7}\left(  \delta\left(  2t\partial_{t}+Y\right)  +K^{3}\right)  =2\delta
t\mathcal{H}-8\delta e^{2K\phi}a^{2}\dot{a}-I^{3}\left(  K^{3}\right). 
\]

Finally, for the third function form $F_{C}\left(  \phi\right)  $, despite the
existence of more elements in the Homothetic algebra of the minisuperspace
there are not any other functional forms of the potential function $V\left(
\phi,\psi\right)  $ where additional Noether symmetries exist.

A natural question that occurs from the above analysis is, whether we can conclude
about the integrability properties of one of the above potential functions or
if we can define invariant functions such that to determine exact solutions.
The dependent variables define a three-dimensional space in which the
evolution of the physical variables takes place. Hence, a specific cosmological model will be called Liouville integrable if there exist at least three
conservation laws which are independent and are in involution. Easily from the
previous results, it follows that this is true for the constant potential
$V_{6}\left(  \phi,\psi\right)  =V_{0}$ and for the exponential potential
$V_{7}\left(  \phi,\psi\right)  =V_{0}e^{-4\delta\phi}$, where at least the
conservations laws $\left\{  \mathcal{H},I^{1}\left(  K^{1}\right)
,I^{3}\left(  K^{3}\right)  \right\},$ and $\left\{  \mathcal{H}%
,I^{1}\left(  K^{1}\right),I^{7}\left(  \delta\left(  2t\partial
_{t}+Y\right)  +K^{3}\right)  \right\}  $ are independent and in involution,
assuming the constraint condition (\ref{dd.04}), $\mathcal{H}\equiv0$.

\section{Exact and analytic solutions}

\label{sec4}

In this section, we present some exact closed-form solutions for the field
equations, as well as the analytic solution for the Liouville integrable
cosmological model. Indeed, we consider the exponential coupling function
$F\left(  \phi\right)  =F_{0}e^{2K\phi}$ where without loss of generality we
select parameter $K=1$.

\subsection{Exact solutions}

Consider now the exponential potential $V\left(  \phi,\psi\right)
=V_{0}e^{-\delta_{1}\psi}e^{-\delta_{2}\phi}$. From the previous analysis it
is clear that, for this specific potential function, the cosmological point-like
Lagrangian (\ref{dd.03}) admits as Noether symmetries the vector fields $\left\{
\delta\left(  2t\partial_{t}+Y\right)  +K^{1},K^{1}+\frac{\delta_{2}}%
{\delta_{1}}K^{3}\right\}  $ and any linear combination of these two.
From these vectors, it is easy to construct the invariant functions
\begin{equation}
a\left(  t\right)  =a_{0}t^{a_{1}},~~\phi\left(  t\right)  =\phi_{1}\ln
t~\text{and}~\psi\left(  t\right)  =\psi_{1}\ln t,~
\end{equation}
where $2-4\delta_{2}\phi_{1}-4\delta_{1}\psi_{1}=0$. By replacing these in the field equations, we end with an algebraic system that
gives the following solutions
\begin{align}
a_{1}  &  =\frac{1-\phi_{1}}{3},~\\
V_{0}  &  =0~,\\
\omega &  =-\frac{16\delta_{1}^{2}\left(  \phi_{1}-1\right)  ^{3}%
+3\beta\left(  1-2\delta_{2}\phi_{1}\right)  }{12\delta_{1}^{2}\phi_{1}^{2}}~,
\end{align}
or%
\begin{align}
a_{1}  &  =\frac{2-\phi_{1}}{3},~\\
V_{0}  &  =\frac{\beta\left(  2\delta_{2}\phi_{1}-1\right)  }{8\delta_{1}^{2}%
},~\\
\omega &  =\frac{8\delta_{1}^{2}\left(  2-\phi_{1}\right)  +3\beta\delta
_{2}\left(  1-2\delta_{2}\phi_{1}\right)  }{6\delta_{1}^{2}\phi_{1}}~,
\end{align}
or%
\begin{align}
a_{1}  &  =\frac{\beta\left(  2\delta_{2}\beta_{1}-1\right)  }{16\delta
_{1}^{2}},~\\
V_{0}  &  =\frac{\beta\left(  2\delta_{2}\phi_{1}-1\right)  \left(
16\delta_{1}^{2}\left(  \phi_{1}-1\right)  +\beta\left(  6\delta_{2}\phi
_{1}-3\right)  \right)  }{128\delta_{1}^{4}},~\\
\omega &  =-\frac{\beta\left(  2\delta_{2}-1\right)  \left(  2\delta_{2}%
\phi_{1}-1\right)  }{4\delta_{1}^{2}\phi_{1}}.
\end{align}
We remark that solutions with $\phi_{1}\psi_{1}=0$ are not accepted because in
this case at least one of the scalar fields does not contribute in the
cosmological fluid.

On the other hand, when $\delta_{1}=0$, that is $\phi_{1}=\frac{1}{2\delta
_{2}}$, we end with the exact solution%
\begin{equation}
a_{1}=\frac{1}{3}\left(  1-\frac{1}{2\delta_{2}}\right)  ,~V_{0}=0\text{ and
}\omega=-\frac{4}{3}\left(  1+\delta_{2}\left(  \delta_{2}\left(  4+3\beta
\psi_{1}^{2}\right)  -4\right)  \right). 
\end{equation}
This is a scaling solution, which means that the cosmological fluid is
described by an ideal gas with the equation of state parameter $w_{\rm eff}$, where
$w_{\rm eff}=\frac{2}{3a_{1}}-1$, or $w_{\rm eff}=\frac{2\delta_{1}+1}{2\delta_{1}-1}%
$. $\ $Thus, acceleration occurs when $-\frac{1}{4}<\delta_{1}<\frac{1}{2}$.

\subsection{Analytic solution for vanishing potential}

For the zero potential function $V\left(  \phi,\psi\right)  =0,~$we define the
new dependent variable $u=3\ln a+\phi$ where the point-like Lagrangian
(\ref{dd.03}) reads%
\begin{equation}
\mathcal{L}\left(  u,\dot{u},\phi,\dot{\phi},\psi,\dot{\psi}\right)
=e^{u+\phi}\left(  \left(  4+3\omega\right)  \dot{\phi}^{2}-8\dot{\phi}\dot
{u}+4\dot{u}^{2}+3\beta\dot{\psi}^{2}\right). 
\end{equation}

We define the canonical momentum components%
\begin{equation}
p_{u}=\frac{\partial\mathcal{L}}{\partial\dot{u}},~p_{\phi}=\frac
{\partial\mathcal{L}}{\partial\dot{\phi}},~p_{\psi}=\frac{\partial
\mathcal{L}}{\partial\dot{\psi}}~,
\end{equation}
that is%
\begin{equation}
\dot{u}=\frac{1}{4\omega}e^{-u-\phi}\left(  \left(  4+3\omega\right)
p_{u}+4p_{\phi}\right)  ~, \label{fe.01}%
\end{equation}%
\begin{equation}
\dot{\phi}=\frac{1}{\omega}e^{-u-\phi}\left(  p_{u}+p_{\phi}\right)  ~,
\label{fe.02}%
\end{equation}%
\begin{equation}
\dot{\psi}=\frac{1}{\beta}e^{-u-\phi}p_{\psi}. \label{fe.03}%
\end{equation}
Hence, we can write the Hamiltonian as follows%
\begin{equation}
\mathcal{H}\equiv\frac{1}{8\omega\beta}e^{-u-\phi}\left(  4\left(
p_{u}+p_{\phi}\right)  ^{2}\beta+4\omega p_{\psi}^{2}+3p_{u}^{2}\omega
\beta\right)  =0. \label{fe.04}%
\end{equation}
The field equations are (\ref{fe.01}), (\ref{fe.02}), (\ref{fe.03})
and
\begin{equation}
\dot{p}_{u}=0~,\dot{p}_{\phi}=0,~\dot{p}_{\psi}=0.
\end{equation}
Hence $p_{u},~p_{\phi}$ and $p_{\psi}$ are integration constants. Consequently, the field equations are of the form%
\[
\dot{u}=u_{1}e^{-u-\phi},~\dot{\phi}=\phi_{1}e^{-u-\phi}\text{~, ~}\dot{\psi
}=\psi_{1}e^{-u-\phi}.
\]

In order to write the analytic solution in closed-form expression, we perform
the change of variables $\frac{{\rm d}u}{{\rm d}t}=\frac{{\rm d}u}{{\rm d}\tau}\frac{{\rm d}\tau}{{\rm d}t}$, with
${\rm d}\tau=e^{-u-\phi}{\rm d}t$ or ${\rm d}t=e^{u+\phi}{\rm d}\tau$. Hence, the field equations are%
\begin{equation}
\frac{{\rm d}u}{{\rm d}\tau}=u_{1},~\frac{{\rm d}\phi}{{\rm d}\tau}=\phi_{1}\text{ and }\frac{{\rm d}\psi
}{{\rm d}\tau}=\psi_{1},
\end{equation}
that is%
\[
u\left(  t\right)  =u_{1}\tau+u_{0},~\phi=\phi_{1}\tau+\phi_{0}\text{ and
}\psi=\psi_{1}\tau+\psi_{0}\text{.}%
\]
Hence, the line element for the physical space is of the form%
\begin{equation}
{\rm d}s^{2}=-e^{2\left(  u_{1}+\phi_{1}\right)  \tau}{\rm d}\tau+e^{\frac{2}{3}\left(
u_{1}-\phi_{1}\right)  t}\left(  {\rm d}x^{2}+{\rm d}y^{2}+{\rm d}z^{2}\right),
\end{equation}
Moreover, we derive the relation $t=\frac{1}{\left(  u_{1}+\phi_{1}\right)
}e^{\left(  u_{1}+\phi_{1}\right)  \tau}$ ,or $\tau=\frac{1}{\left(
u_{1}+\phi_{1}\right)  }\ln\left(  \left(  u_{1}+\phi_{1}\right)  t\right)  $.
Thus, the line element for the FLRW spacetime in the lapse function $N\left(
t\right)  =1$, becomes
\begin{equation}
{\rm d}s^{2}=-{\rm d}t+\left(  \left(  u_{1}+\phi_{1}\right)  t\right)  ^{\frac{2}{3}%
\frac{\left(  u_{1}-\phi_{1}\right)  }{\left(  u_{1}+\phi_{1}\right)  }%
}\left(  {\rm d}x^{2}+{\rm d}y^{2}+{\rm d}z^{2}\right). 
\end{equation}

The latter solution describe an ideal gas solution with the constant equation of
state parameter $w_{\rm eff}=-1+2\frac{u_{1}+\phi_{1}}{u_{1}-\phi_{1}}~$, from
where we infer that the solution describes an accelerated universe
when $\frac{u_{1}+\phi_{1}}{u_{1}-\phi_{1}}<\frac{1}{3}$.

\subsection{Analytic solution for constant potential}

Consider now the constant potential $V\left(  \phi,\psi\right)  =V_{0}$. The solution process is similar to before. We define the new variable $U=3\ln
a+2\phi$, and in terms of the Hamiltonian formalism, the field equations are%
\begin{equation}
\dot{U}=\frac{1}{4\omega}e^{-U}\left(  \left(  16+3\omega\right)
p_{U}+8p_{\phi}\right)  ~,
\end{equation}%
\[
\dot{\phi}=\frac{1}{\omega}e^{-U}\left(  2p_{U}+p_{\phi}\right)  ~,
\]%
\[
\dot{\psi}=\frac{1}{\beta}e^{-U}p_{\psi},
\]
and%
\begin{equation}
\dot{p}_{U}=\frac{1}{4\beta\omega}e^{-U}\left(  4\left(  2p_{U}+p_{\phi
}\right)  ^{2}\beta+4\omega p_{\psi}^{2}+3\beta\omega p_{U}^{2}\right)  ~,
\end{equation}%
\begin{equation}
\dot{p}_{\phi}=0,~\dot{p}_{\psi}=0~,
\end{equation}
where the constraint equation is
\begin{equation}
\left(  16+3\omega\right)  p_{U}^{2}+\frac{4\omega}{\beta}p_{\psi}^{2}%
+16p_{U}p_{\phi}+4\left(  p_{\phi}^{2}-2e^{2U}\omega V_{0}\right)  =0\text{.}%
\end{equation}
Consequently, the conservation laws for the field equations are the momentum
$p_{\phi}$ and $p_{\psi}$.

We proceed with the derivation of the Action $S\left(  U,\phi,\psi\right)  $
by solving the Hamilton-Jacobi equation%
\begin{equation}
\left(  16+3\omega\right)  \left(  \frac{\partial S}{\partial U}\right)
^{2}+\frac{4\omega}{\beta}\left(  \frac{\partial S}{\partial\psi}\right)
^{2}+16\left(  \frac{\partial S}{\partial U}\right)  \left(  \frac{\partial
S}{\partial\phi}\right)  +4\left(  \left(  \frac{\partial S}{\partial\phi
}\right)  ^{2}-2e^{2U}\omega V_{0}\right)  =0\text{,}%
\end{equation}
where the conservation laws give $\left(  \frac{\partial S}{\partial\phi
}\right)  -p_{\phi0}=0$ and $\left(  \frac{\partial S}{\partial\psi}\right)
-p_{\psi0}=0$. Thus, it follows
\[
S\left(  U,\phi,\psi\right)  =S_{1}\left(  U\right)  +p_{\phi0}\phi+p_{\psi
0}\psi,
\]
where $S_{1}\left(  U\right)  $ is given by the first-order ordinary
differential equation%
\begin{equation}
\frac{dS_{1}\left(  U\right)  }{dU}=\frac{2\sqrt{\omega\left(  \left(
16+3\omega\right)  \left(  2e^{2A}V_{0}\beta-p_{\psi0}^{2}\right)  -3\beta
p_{\phi0}^{2}\right)  }-8\sqrt{\beta}p_{\phi0}}{\sqrt{\beta}\left(
16+3\omega\right)  }\text{.}%
\end{equation}
Therefore, the field equations are reduced in the following system%
\begin{equation}
\dot{U}=\frac{1}{4\omega}e^{-U}\left(  \left(  16+3\omega\right)  \frac
{dS_{1}\left(  U\right)  }{dU}+8p_{\phi0}\right)  ~,
\end{equation}%
\begin{equation}
\dot{\phi}=\frac{1}{\omega}e^{-U}\left(  2\frac{dS_{1}\left(  U\right)  }%
{dU}+p_{\phi0}\right)  ~,
\end{equation}%
\begin{equation}
\dot{\psi}=\frac{1}{\beta}e^{-U}p_{\psi0}.
\end{equation}
The analytic solution is expressed as follows%
\begin{equation}
e^{U\left(  t\right)  }=\frac{\left(  \exp\left(  \sqrt{\frac{V_{0}\left(
16+3\omega\right)  }{2\omega}}t\right)  +2V_{0}\beta\left(  16+3\omega\right)
\left(  3p_{\phi0}^{2}\beta+p_{\psi0}^{2}\left(  16+3\omega\right)  \right)
\exp\left(  -\sqrt{\frac{V_{0}\left(  16+3\omega\right)  }{2\omega}}t\right)
\right)  }{4V_{0}\beta\left(  16+3\omega\right)  },
\end{equation}%
\begin{align}
\phi\left(  t\right)   &  =\phi_{1}+\frac{2\sqrt{\omega}}{\sqrt{V_{0}%
16+3\omega}}U\left(  t\right) \nonumber\\
&  +\frac{4\left(  3p_{\phi0}\sqrt{\omega\beta}\right)  }{\left(
16+3\omega\right)  \sqrt{3\beta p_{\phi0}^{2}+p_{\psi0}^{2}\left(
16+3\omega\right)  }}\arctan\left(  \frac{\exp\left(  \sqrt{\frac{V_{0}\left(
16+3\omega\right)  }{2\omega}}t\right)  }{\sqrt{2V_{0}\beta\left(
16+3\omega\right)  \left(  3\beta p_{\phi0}^{2}+p_{\psi0}^{2}\left(
16+3\omega\right)  \right)  }}\right)  ~,
\end{align}
and%
\begin{equation}
\psi\left(  t\right)  =\frac{p_{\psi0}}{\beta}\int e^{-U}dt\text{~}.
\end{equation}

We conclude that in order for the solution to be real, the following constraints
follow $\frac{V_{0}\left(  16+3\omega\right)  }{2\omega}>0$, and $3\beta
p_{\phi0}^{2}+p_{\psi0}^{2}\left(  16+3\omega\right)  >0$. In the late universe, the asymptotic behaviour of the analytic solution is
\[
e^{U\left(  t\right)  }\simeq\exp\left(  \sqrt{\frac{V_{0}\left(
16+3\omega\right)  }{2\omega}}t\right), \phi\left(  t\right)  \simeq
U\left(  t\right). 
\]
that is, $a\simeq e^{a_{1}t},$ from where it follows that the de Sitter
Universe describes the late-time evolution of this specific cosmological model.

\subsection{Analytic solution for exponential potential}

In the case of the exponential potential $V\left(  \phi\right)  =V_{0}%
e^{-4\delta\phi}$ we proceed with the definition of the new dependent variable
$A=3\ln a+2\left(  1-\delta\right)  \phi$. The field equations are
written in the Hamiltonian formalism%
\begin{equation}
e^{A+2\phi}\dot{A}=\frac{1}{4\omega}\left(  3\omega p_{A}+8\left(  p_{\phi
}+2p_{A}\left(  1-\delta\right)  \right)  \left(  1-\delta\right)  \right)  ~,
\end{equation}%
\[
e^{A+2\phi}\dot{\phi}=\frac{1}{\omega}\left(  p_{\phi}+2\left(  1-\delta
\right)  p_{A}\right)  ,~
\]%
\begin{equation}
e^{A+2\phi}\dot{\psi}=\frac{1}{\beta}p_{\psi}~,
\end{equation}%
\begin{equation}
e^{A+2\phi}\dot{p}_{A}=\frac{1}{4\omega\beta}\left(  4\beta\left(  p_{\phi
}+2p_{A}\left(  1-\delta\right)  \right)  ^{2}+4\omega p_{\psi}^{2}%
+3\beta\omega p_{A}^{2}\right)  ~,
\end{equation}%
\[
\dot{p}_{\phi}=0,~\dot{p}_{\psi}=0.
\]
with constraint equation%
\begin{equation}
4p_{\psi}^{2}\omega+\beta\left(  16p_{A}p_{\phi}\left(  1-\delta\right)
+p_{A}^{2}\left(  16\left(  1-\delta\right)  ^{2}+3\omega\right)  +4\left(
p_{\phi}^{2}-2V_{0}\omega e^{2A}\right)  \right)  =0.
\end{equation}
Consequently, $p_{\phi}\left(  t\right)  =p_{\phi0}$ and $p_{\psi}\left(
t\right)  =p_{\psi0}$ are the two conservation laws.

We write the Hamilton-Jacobi equation and with the use of the conservation
laws, we derive the following functional form for the Action%
\begin{equation}
S\left(  A,\phi,\psi\right)  =S_{1}\left(  A\right)  +p_{\phi0}\phi+p_{\psi
0}\psi,
\end{equation}
where now%
\begin{equation}
\frac{dS_{1}\left(  A\right)  }{dS}=\frac{8p_{\phi0}\sqrt{\beta}\left(
1-\delta\right)  +2\sqrt{\omega\left(  \left(  2e^{2A}\beta V_{0}-p_{\psi
0}^{2}\right)  \left(  16\left(  1-\delta\right)  ^{2}+3\omega\right)  -3\beta
p_{\phi0}^{2}\right)  }}{\sqrt{\beta}\left(  16\left(  1-\delta\right)
^{2}+3\omega\right)  },
\end{equation}
Hence, by replacing with $p_{A}=\frac{dS_{1}\left(  A\right)  }{dS}$ we end
with a system of three first-order differential equations.

We can now derive the scalar fields $\phi,~\psi$ as function of $A$,
that is,
\begin{align}
\phi\left(  A\right)   &  =\phi_{1}-\frac{8\left(  1-\delta\right)
}{16\left(  1-\delta\right)  ^{2}+3\omega}A\nonumber\\
&  -\frac{6p_{\phi0}\sqrt{\omega\beta}\arctan\left(  \frac{\sqrt{2e^{A}\beta
V_{0}\left(  16\left(  1-\delta\right)  ^{2}+3\omega\right)  -\left(
3p_{\phi0}^{2}\beta+p_{\psi0}^{2}\left(  16\left(  1-\delta\right)
^{2}+3\omega\right)  \right)  }}{\sqrt{3p_{\phi0}^{2}\beta+p_{\psi0}%
^{2}\left(  16\left(  1-\delta\right)  ^{2}+3\omega\right)  }}\right)
}{\left(  16\left(  1-\delta\right)  ^{2}+3\omega\right)  \left(
\sqrt{3p_{\phi0}^{2}\beta+p_{\psi0}^{2}\left(  16\left(  1-\delta\right)
^{2}+3\omega\right)  }\right)  },
\end{align}%
\begin{equation}
\psi\left(  A\right)  =\frac{2p_{\psi0}^{2}\sqrt{\omega}\arctan\left(
\frac{\sqrt{2e^{A}\beta V_{0}\left(  16\left(  1-\delta\right)  ^{2}%
+3\omega\right)  -\left(  3p_{\phi0}^{2}\beta+p_{\psi0}^{2}\left(  16\left(
1-\delta\right)  ^{2}+3\omega\right)  \right)  }}{\sqrt{3p_{\phi0}^{2}%
\beta+p_{\psi0}^{2}\left(  16\left(  1-\delta\right)  ^{2}+3\omega\right)  }%
}\right)  }{\sqrt{\beta\left(  3p_{\phi0}^{2}\beta+p_{\psi0}^{2}\left(
16\left(  1-\delta\right)  ^{2}+3\omega\right)  \right)  }}\text{. }%
\end{equation}

In order to write the analytic solution in closed-form expression in terms of
the independent variable, we make the change of variable $dt=e^{A+2\phi}d\tau
$.\ Thus, in terms of $\tau$ the analytic solution reads%
\begin{equation}
e^{A\left(  \tau\right)  }=\sqrt{\frac{3p_{\phi0}^{2}\beta+p_{\psi0}%
^{2}\left(  16\left(  1-\delta\right)  ^{2}+3\omega\right)  }{2\beta
V_{0}\left(  16\left(  1-\delta\right)  ^{2}+3\omega\right)  }}\left(
\cos\left(  \frac{1}{2}\sqrt{\frac{3p_{\phi0}^{2}\beta+p_{\psi0}^{2}\left(
16\left(  1-\delta\right)  ^{2}+3\omega\right)  }{\beta\omega}}\tau\right)
\right)  ^{-2}%
\end{equation}%
\[
\phi\left(  \tau\right)  =\phi_{1}+\frac{3p_{\phi0}\sqrt{\omega\beta}%
-8\sqrt{\beta}\left(  1-\delta\right)  \sqrt{\omega}\ln\left(  \cos\left(
\frac{1}{2}\sqrt{\frac{3p_{\phi0}^{2}\beta+p_{\psi0}^{2}\left(  16\left(
1-\delta\right)  ^{2}+3\omega\right)  }{\beta\omega}}\tau\right)  \right)
}{2\sqrt{\omega\beta}\left(  16\left(  1-\delta\right)  ^{2}+3\omega\right)
}~,
\]
and%
\[
\psi\left(  \tau\right)  =-\frac{p_{\psi0}}{\beta}\tau\text{.}%
\]
It is easy to see that the de Sitter universe can be recovered as a late-time attractor.

\section{Conclusions}
\label{conc}

In this work, we considered a two scalar field cosmology, where one of the fields couples with a dilatonic coupling to gravity, while the second one couples minimally; there is, however, an interacting potential between the two fields. As explained earlier, there is a strong motivation for using multiple scalar fields in cosmology since it can provide much richer phenomenology with both inflationary and late-time dark energy models. In curvature gravity, many studies have considered more than one scalar field, but in the teleparallel geometry, where torsion is responsible for the gravitational forces instead of curvature, the works performed are limited.

We performed a symmetry analysis to classify those models that are invariant under point transformations. We find seven classes of the coupling function and the interaction potential for which non-trivial conservation laws exist. The results are summarized in Table~\ref{tab:results-symmetries}. In the first column, we have the form of the coupling function $F(\phi)$; in the second one, the form of the potential $V(\phi,\psi)$; in the third one, the symmetry vector and in the last one the corresponding conservation law.
\begin{table}[h!]\footnotesize
\begin{tabular}{|c|c|c|}%
\hline
 Potential $V(\phi,\psi)$ & Symmetry vector & Conservation Law \\
\hline\hline
$V(\phi)$ & $K^1=\partial _\psi$ & $I^1 = a^3 e^{2K\phi} \beta \dot{\psi}$\\
\hline
 $e^{-4\delta \psi }V(\phi)$ & $\delta (2t\partial _t + Y) +K^1$ & $I^2 = 2 \delta t \mathcal{H} - e^{2K\phi} a^2 (8 \delta \dot{a} + \beta a \dot{\psi})$\\
\hline
$V(\psi)$ & $K^3 = - \frac{2}{3}Ka\partial _a + \partial _\phi$ & $I^3 = -e^{2K\phi} a^2 (8 K \dot{a}-\omega a \dot{\phi})$\\
\hline
 $V(\psi - \frac{\phi}{\alpha})$ & $K^1 + \alpha K^3 $ & $I^4 = I^1 + \frac{1}{\alpha}I^3$\\
\hline
 $V(\alpha\psi - \phi)e^{-\frac{\delta}{a}\psi}$ & $\delta (2t\partial _t + Y) +K^1 + \alpha K^3 $ & $I^5 = I^2 (\delta (2t\partial _t + Y)+K^1) + \alpha I^3$\\
\hline
$V_0$ & $K^2 = -\frac{2}{3}K a \psi \partial _a + \psi \partial _\phi + \frac{8K \ln a - \omega \phi}{\beta}\partial _\psi $ & $I^6 = e^{2K\phi} a^2\left( -8Ka \psi \dot{a}+ \omega a \psi \dot{\phi} + a \dot{\psi}(8K\ln a - \omega \phi)\right) $ \\
\hline
$V(\psi) e^{-4 \delta \phi}$ & $\delta (2t\partial _t + Y)+K^3 $ & $I^7 = 2 \delta t \mathcal{H} - 8 \delta e^{2K\phi} a^2 \dot{a} - I^3$ \\
\hline
\end{tabular}
\caption{Classification according to the Noether Symmetry analysis. In the first two cases, the coupling function is an arbitrary $F(\phi)$, and in the rest, it is $F_B(\phi) = F_0 e^{2K\phi}$ with $K$ being a constant. The proper Homothetic vector $Y$ is given by Eq.~\eqref{eq:HomVec}.}\label{tab:results-symmetries}
\end{table}

The symmetry analysis would mean nothing if it could not help us find exact solutions for the system. We have shown that when the coupling function is of the form $F(\phi) = F_0 e^{2 \phi}$, we can find some closed-form analytic solutions for the scale factor and the scalar fields for four different configurations of the interacting potential $V(\phi,\psi)$. In a follow-up work, we plan to study the stability of these solutions. 

\bigskip

\section*{Acknowledgments} \label{sec10}
The work was supported by Nazarbayev University Faculty Development Competitive Research Grant No. 11022021FD2926 and by the Hellenic Foundation for Research and Innovation (H.F.R.I.) under the “First Call for H.F.R.I. Research Projects to support Faculty members and Researchers and the procurement of high-cost research equipment grant” (Project Number: 2251). This article is based upon work from COST Action CA21136 Addressing observational tensions in cosmology with systematics and fundamental physics (CosmoVerse) supported by COST (European Cooperation in Science and Technology).
GL was funded by Vicerrectoría de Investigación y Desarrollo Tecnológico (Vridt) at Universidad Católica del Norte through Concurso De Pasantías De Investigación Año 2022, Resolución Vridt No. 040/2022 and through Resolución Vridt No. 054/2022. He also thanks the support of Núcleo de Investigación Geometría Diferencial y Aplicaciones, Resolución Vridt No. 096/2022. AP acknowledges Vridt-UCN through Concurso de Estadías de Investigación, Resolución VRIDT N°098/2022.

\end{document}